\documentclass[aps,prl,twocolumn]{revtex4}

\usepackage{amssymb}
\usepackage{graphicx}
\usepackage{epstopdf}
\usepackage{amsmath}
\usepackage{epsfig}
\usepackage{color}
\usepackage{ulem}

\begin{document}
\title{Effective negative specific heat by destabilization of metastable states in dipolar systems}

\author{Vazha Loladze$^{1,2}$, Thierry Dauxois$^{3}$, Ramaz Khomeriki$^{1}$, Stefano Ruffo$^{4,5,6}$ }
\affiliation {${\ }^1$Physics Department, Javakhishvili Tbilisi
State University, Chavchavadze 3, 0128 Tbilisi, Georgia \\
${\ }^2$Department of Physics, Florida State University, Tallahassee, FL 32306, USA \\
${\ }^3$Univ. Lyon, ENS de Lyon, Univ. Claude Bernard, CNRS,
Laboratoire de Physique, F-69342 Lyon, France \\
${\ }^4$SISSA, Via Bonomea 265, I-34136 Trieste, Italy \\
${\ }^5$INFN, Sezione di Trieste, I-34151 Trieste, Italy\\
${\ }^6$Istituto dei Sistemi Complessi, Consiglio Nazionale delle
Ricerche, via Madonna del Piano 10, I-50019 Sesto Fiorentino, Italy}

\begin{abstract}
We study dipolarly coupled three dimensional spin systems in both the microcanonical and the canonical ensembles by introducing appropriate numerical methods to determine the microcanonical temperature and by realizing a canonical model of heat bath. In the microcanonical ensemble, we show the existence of a branch of stable antiferromagnetic states in the low energy region. Other metastable ferromagnetic states exist in this region: by externally perturbing them, an effective negative specific heat is obtained. In the canonical ensemble, for low temperatures, the same metastable states are unstable and reach a new branch of more robust metastable states which is distinct from the stable one. Our statistical physics approach allows us to put some order in the complex structure of stable and metastable states of dipolar systems.

\end{abstract}
 \maketitle

For long-range interactions~\cite{bookLRI,levin}, the dimensionality of the sample is larger or equal than the decay exponent of the power law interaction itself. Among the most exotic manifestations of long-range interactions  is negative specific heat. As an example, for self gravitating systems there is an energy region where  temperature increases with a loss of energy~\cite{antonov,lynden,thirring}.  Experimental observations of negative specific heat have been reported for atomic clusters near phase transitions~\cite{solid,gas}. On the theoretical side, two-dimensional systems~\cite{bouchet} and mean-field models~\cite{bookLRI} have been shown to display negative specific heat in the microcanonical ensemble. On the other hand, it has been observed that the phenomenon of negative specific heat is tightly related with the general notion of inequivalence of ensembles~\cite{campa1,PRL1}. However, there is not yet a clear laboratory experiment which shows negative specific heat for samples of macroscopic size, where one could directly study the phenomenon and devise possible applications.

\begin{figure}[b]
\center
\includegraphics[scale=.33]{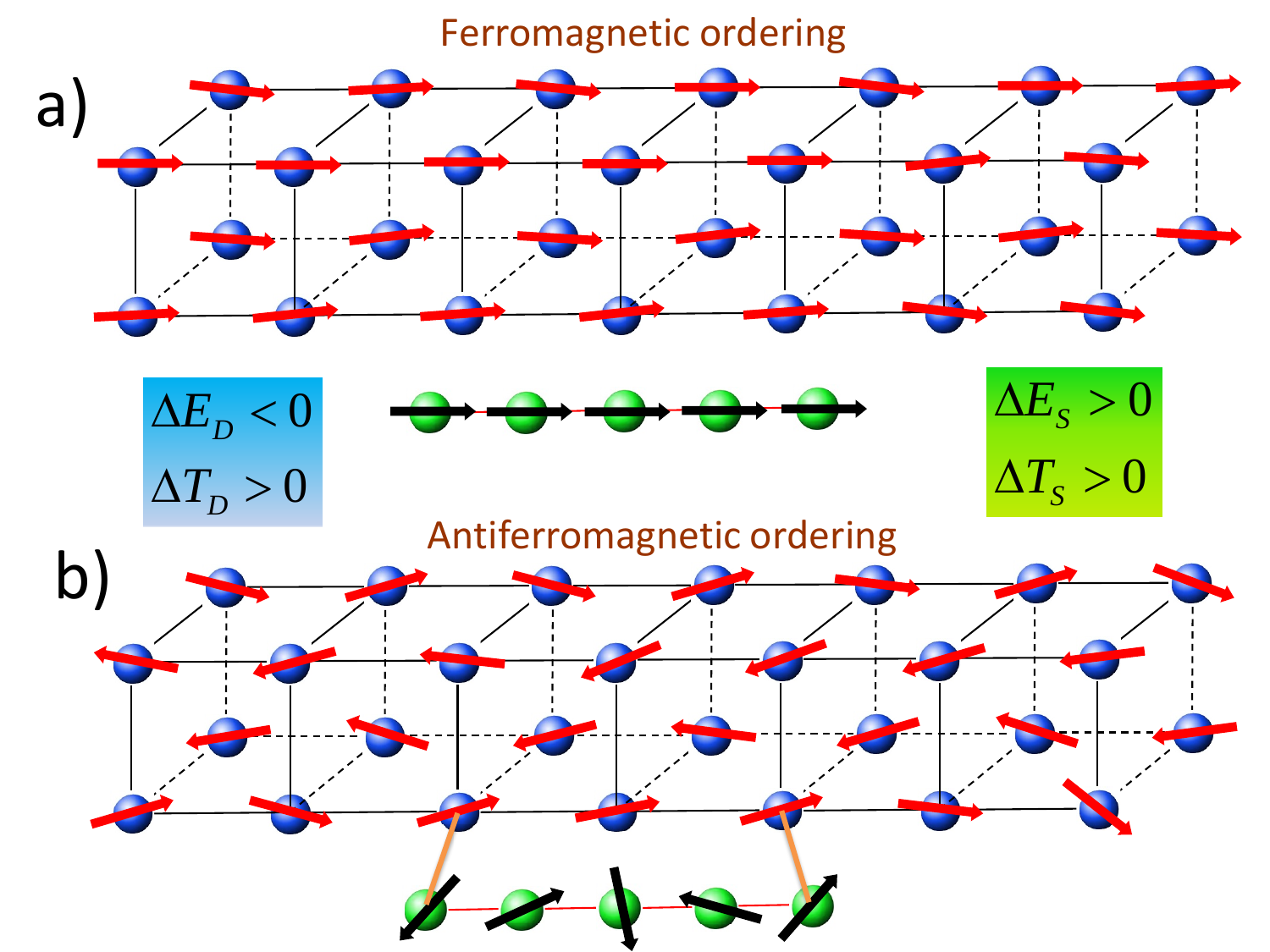}
\caption{Schematic representation of a dipolar spin system~(D) coupled to a short-range spin chain (S): a) initial state where both D and S are ferromagnetic; b) final state where D is antiferromagnetic and S is paramagnetic. In the transition, $\Delta E_D<0$ and $\Delta T_D>0$ showing a negative specific heat of~D, while the specific heat of the bath is positive: $\Delta E_S>0$ and $\Delta T_S>0$.}
\label{ferpar0}
\end{figure} 
Dipolar forces are marginal examples among long-range interacting systems,
because the interaction strength among spins decays as $1/r^3$ in three dimensions.  There is an open question whether such forces could induce negative specific heat and ensemble inequivalence. Previously, another feature of long-range interactions, i.e. ergodicity breaking, was shown for a model of dipolar spins by reducing it to an effective mean-field model~\cite{mukamel,miloshevich}. Numerical simulations show that this exotic feature appears only for needle shaped samples (i.e. when the aspect ratio is large) and when, additionally, one has spontaneous magnetization in zero field. However, it has been argued in Ref. \cite{miloshevich} that, in the thermodynamic limit \cite{griffiths}, the ferromagnetic state survives only in the case of a body-centered cubic lattice. Despite this fact, we show in the present paper that, when considering a finite simple cubic lattice with $2\times 2$ base and a large aspect ratio ergodicity breaking and negative specific heat are found. By slightly increasing the base size to $3\times 3$, the sample looses all these features, which are therefore absent also in the thermodynamic limit. However, we show below that the simple cubic lattice displays also a ferrimagnetic (only partially magnetized) state which could persist also in the thermodynamic limit and show some of the exotic properties of long-range interactions.  

Realizing the microcanonical ensemble for spin systems is straightforward by the direct integration of the Hamiltonian equations of motion. On the contrary, the realization of canonical ensemble is more complex. It requires the introduction of the Landau-Gilbert damping term in the equations of motion and the coupling with a Nos\'e-Hoover thermostat adapted to spin variables~\cite{term}. 

In this Letter, we consider dipolarly coupled spin systems and we study them in both the microcanonical and canonical ensembles. These systems could be experimentally realized using e.g. cobalt nanoparticles~\cite{cobalt}.
In particular, we consider a setup, in which a magnetized dipolar system with a simple cubic lattice structure interacts with a short-range spin chain (see Fig.~\ref{ferpar0}). Due to the interaction, the magnetized state is destabilized and converts
into an antiferromagnetic one. If one defines the specific heat of the system in this particular process as the ratio between the exchanged energy and the microcanonical temperature difference, the effective specific heat takes negative values depending on the size of the sample. The same process, when observed in the canonical ensemble, follows a completely different pathway which leads in some cases to a metastable partially ferromagnetic state. 

The Hamiltonian of our system consists of classical spins interacting through dipolar forces and can be written as follows
\begin{eqnarray}
{\cal  H} = \frac{\epsilon}{2}\sum_{i\neq j}\frac{a^3}{r_{ij}^3}
\left({\bf S}_i\cdot{\bf S}_j-3\frac{({\bf S}_i\cdot
{\bf  r}_{ij})({\bf S}_j\cdot {\bf r}_{ij})}{r_{ij}^2}\right),
\label{1}
\end{eqnarray}
where ${\bf S}_i$ is a unit spin vector attached to the $i$-th site of a lattice and ${\bf r}_{ij}$ is a displacement vector between the $i$-th and the $j$-th site. In this problem, we have a length scale~$a$ (lattice spacing) and an energy scale $\epsilon=\mu_0\sigma^2/(4\pi a^3)$ of the dipolar interactions. For instance, in the case of cobalt nanoparticles with magnetic moment $\sigma\sim 2\cdot 10^5\mu_B$ and separation length $a\sim 20$~nm (see e.g. Ref. \cite{cobalt}), this energy could be as large as $2500$~K ($\mu_0$ is the vacuum permeability and $\mu_B$ the Bohr magneton). The dynamics of the system is described by the equations of motion
\begin{eqnarray}
\frac{d{\bf S}_i}{dt}= {\gamma}\left[{\bf S}_i\times{\bf H}_i\right]
\quad \mbox{where} \quad {\bf H}_i=-\frac{1}{\sigma}\frac{\partial
{\cal H}}{\partial {\bf S}_i}.
\label{2}
\end{eqnarray}
Here, $\gamma$ is the gyromagnetic ratio, distance is measured in units of $a$ and time in units of $t_0=2\sigma/(\gamma\epsilon)$. We are considering dipolar needles with simple cubic lattice structure that consists of 200 spins with geometry $2\times2\times50$. If one restricts the interaction between the spins only to nearest neighbors in Eq.~\eqref{1}, then the ground state is antiferromagnetic in the transverse direction and ferromagnetic in the longitudinal direction: there cannot be any magnetized state in the system. However, if the interaction is all-to-all, long-range effects cause the appearance of stable magnetized states in the microcanonical ensemble.

In Ref.~\cite{miloshevich}, model~\eqref{1} has been mapped into a reduced one-dimensional spin chain where each spin in the chain represents the averaged magnetization in a transversal layer of the original model. Then, the Hamiltonian of the reduced one-dimensional model describes competing antiferromagnetic short-range and ferromagnetic mean-field interactions. The reduced model can be solved analytically in both the microcanonical and the canonical ensembles. The phase diagram reveals ensemble inequivalence with negative specific heat in the microcanonical case and temperature jumps~\cite{mukruf}. Although the correspondence of the initial three-dimensional model to its reduced one-dimensional counterpart is qualitatively well justified, the quantitative description is not satisfactory and one cannot directly relate the analytical results for one-dimensional model to the realistic three-dimensional situation~\eqref{1}. Therefore, one should find ways to deal directly with the three-dimensional Hamiltonian using numerical simulations.

As a first step in this direction, we recall the definition of microcanonical temperature $1/T_\mu=\partial {\cal S}/\partial E$  as the ratio between an infinitesimal change of energy and the corresponding shift in entropy~\cite{rugh,gut}. From this definition follows a well known {\it caloric curve} in the case of one-dimensional spin chains with nearest neighbor
coupling $K$ 
\begin{equation}
 T_\mu=\frac{K}{\mathcal{L}^{-1}(-{\cal E}/K)},
\label{lang}
 \end{equation}
where ${\cal E}$ is the energy per spin and $\mathcal{L}(x)\equiv \coth(x)-1/x$ is the Langevin function. The same microcanonical formula for
the temperature can be used for the dipolar Hamiltonian~\eqref{1} and one obtains the explicit expression~\cite{wira} 
\begin{equation}
\frac{1}{T_\mu}=\sum_{j}\left[{\bf S}_j\times{\bf \nabla}_j\right]\cdot\frac{\left[{\bf S}_j\times{\bf \nabla}_j\right]{\cal H}}{\Bigl| \sum_{i}\left[{\bf S}_i\times {\bf\nabla}_i\right] {\cal H}\Bigr|^2},
\label{temp}
\end{equation}
where ${\bf \nabla}_j\equiv\partial/\partial{\bf S}_j$. This method allows us to compute numerically the instantaneous microcanonical temperature as a dynamical variable.

The next step consists in constructing a thermal bath in order to study our system within the canonical ensemble. For this
purpose, we couple each dipolar spin ${\bf S}$ to a heat bath spin ${\bf J}$ via the following short-range coupling
\begin{equation}
 H_{sh}=-\frac{\epsilon}{2}{\bf S}\cdot{\bf J}. 
 \label{sh}
 \end{equation}
Heat bath spins ${\bf J}$ are thermalized using the method introduced in Ref.~\cite{term} since the standard 
Nos\'e-Hoover thermostat~\cite{Nose} cannot work here because spins are not canonical variables and, moreover, 
one has to keep the spin length fixed. We have checked that even if we couple only partially the bulk dipolar 
spins (in order to save simulation time), we obtain similar results to when we couple all the spins of the bulk. 
According to Ref.~\cite{term}, bath spins are described by adding a spin length preserving damping term to 
the equation of motion~\eqref{1} as follows
 \begin{equation}
 \frac{d{\bf J}}{dt}=\left[{\bf J}\times{\bf S}\right]+\sum_\ell g_\ell(\eta_\ell)\left[{\bf J}\times{\bf A}_\ell\right],
\label{LLG}
\end{equation}
where index $\ell$ numbers the damping terms (typically we use two of them in the thermalization process), 
${\bf A}_\ell$ are  arbitrary vectors satisfying the conditions $\partial/\partial {\bf J}\times {\bf A}_\ell\neq 0$ and $\eta_\ell$ are additional phase space variables. In order to derive the equations of motions of these additional variables, one hase to extend the phase space and define the probability distribution function $f({\bf J},\eta_\ell)\sim\exp\left(-[{\cal H}_{sh}+\sum_\ell G_\ell(\eta_\ell)/\alpha_\ell]/T\right)$ with canonical temperature $T$ and use Liouville continuity equation in terms of the bath spin variables ${\bf J}$ and of the heat bath additional degrees of freedom $\eta_\ell$ with the condition $\partial \dot\eta_\ell/\partial\eta_\ell=0$. Then, using Eq.~\eqref{LLG}, one gets the condition $g_\ell(\eta_\ell)=\partial G_\ell(\eta_\ell)/\partial \eta_\ell$ and the equations of motion for $\eta_\ell$
\begin{equation}
 \frac{d\eta_\ell}{dt}=\alpha_\ell\left\{ {\bf S}\cdot \left[{\bf J}\times{\bf A}_\ell\right]-T\left[\frac{\partial}{\partial {\bf J}}\times{\bf A}_\ell\right]\cdot{\bf J}\right\}.
\label{bath}
\end{equation}
As anticipated, in the numerical simulations, we use two damping terms with the corresponding additional heat bath phase space degrees of freedom $\eta_1$ and $\eta_2$. Moreover, we choose the arbitrary vectors as ${\bf A}_1=(J_z,J_x,J_y)$, ${\bf A}_2=(J_y,-J_x,0)$ and the damping functions $g_1(\eta_1)=(\eta_1)^3$ and $ g_2(\eta_2)=\eta_2$. In all our numerical simulations, we set the heat bath contact values to $\alpha_1=\alpha_2=2$. Thus, the heat bath evolution Eqs.~\eqref{LLG} and~\eqref{bath}, together with Eqs.~\eqref{1},~\eqref{2} and~\eqref{sh} associated with the dipolar spin system, provide a canonical set-up for our dipolar magnetic needle. The microcanonical set-up is simply obtained by removing the heat bath and measuring the temperature via formula~\eqref{temp}.
\begin{figure}[t]
\center
\includegraphics[scale=.5]{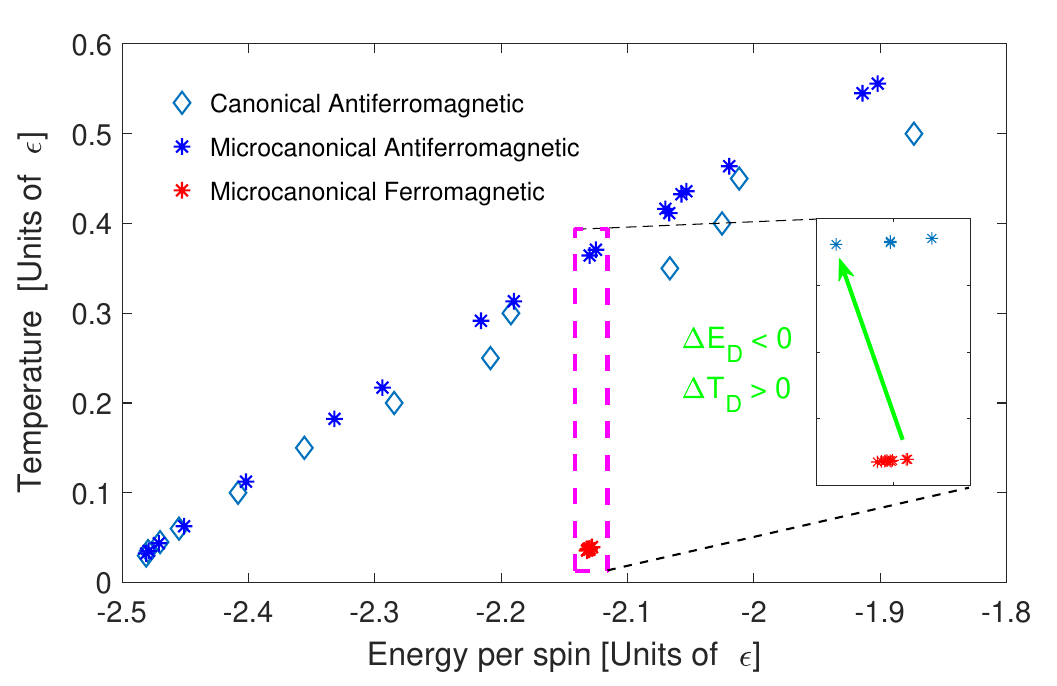}
\caption{{\it Caloric curves} (temperature vs. energy) for different phases in the canonical and microcanonical ensembles. 
The arrow shows a transition from microcanonical ferromagnetic to antiferromagnetic states in which the specific heat  is negative. The transition occurs when the dipolar system is put in contact with a short-range
spin chain which perturbs the metastable ferromagnetic state.} 
\label{ET}
\end{figure}

We have tested the validity of the expression for microcanonical temperature ~\eqref{temp} by using it in conditions of
thermal equilibrium, when the dipolar system is in contact with the canonical thermal bath at temperature $T$: the value of the time averaged dynamical temperature~$T_\mu$ was always in excellent agreement with the thermal bath temperature 
$T$. On the other hand, we have connected a short spin chain with nearest neighbor coupling $K$ (typically $K=\epsilon/2$)
to the dipolar needle with the aim of using it as a ``thermometer'' of the dipolar system: in this case the spin chain
temperature obtained with formula~\eqref{lang} coincides with the microcanonical temperature computed using
formula~\eqref{temp} for the dipolar system. In order to reduce numerical errors and to obtain a better energy conservation, we have devised a second order symplectic integrator~\cite{symplectic} inspired by Yoshida' approach~\cite{yoshida}.

We are now ready to discuss {\it caloric curves} (temperature vs. energy) for both the microcanonical and
canonical ensembles in different conditions.  In microcanonical simulations, we have started the system in 
both a fully magnetized state along the needle longitudinal direction and in an ``antiferromagnetic'' initial
state (antiferromagnetic in the transverse plane and ferromagnetic in the longitudinal direction):
see~Fig.~\ref{ferpar0}. In both cases, in order to increase the energy of the initial state, we add a 
small random transversal component to each spin. Temperature is measured by averaging over time 
the dynamical quantity~\eqref{temp} in the steady state. We show in Fig.~\ref{ET} the results of these
simulations by representing the {\it caloric curves} (temperature vs. energy) of both the antiferromagnetic
and the magnetic states. In agreement with Ref.~\cite{miloshevich}, we see the coexistence of 
antiferromagnetic and magnetic states in a tiny range of energy per spin [~-2.135$\epsilon$,~-2.125$\epsilon$]  if the aspect ratio of the system is quite large. In particular, in order to realise a robust ferromagnetic state one should take at least a $2\times 2\times 32$ dipolar system. Therefore, in numerical simulations, we use a quasi-one-dimensional sample of the size $2\times  2 \times 50$ in order to clearly monitor the phase transition process. 
For the magnetized states, we monitor approximately the same averaged magnetization $\sim 0.94$.
Figure~\ref{ET} also shows that there is a sharp temperature jump between antiferromagnetic
and ferromagnetic states in the range where the energies of those states are the same when the 
dipolar system is put in contact with a short-range spin chain which destabilizes the ferromagnetic
state. It is also clearly seen that the jump is accompanied by a small energy decrease (see the arrow in 
the inset of Fig.~\ref{ET}) which induces a negative specific heat~$C_D < 0$.
\begin{figure}[b]
\includegraphics[scale=.5]{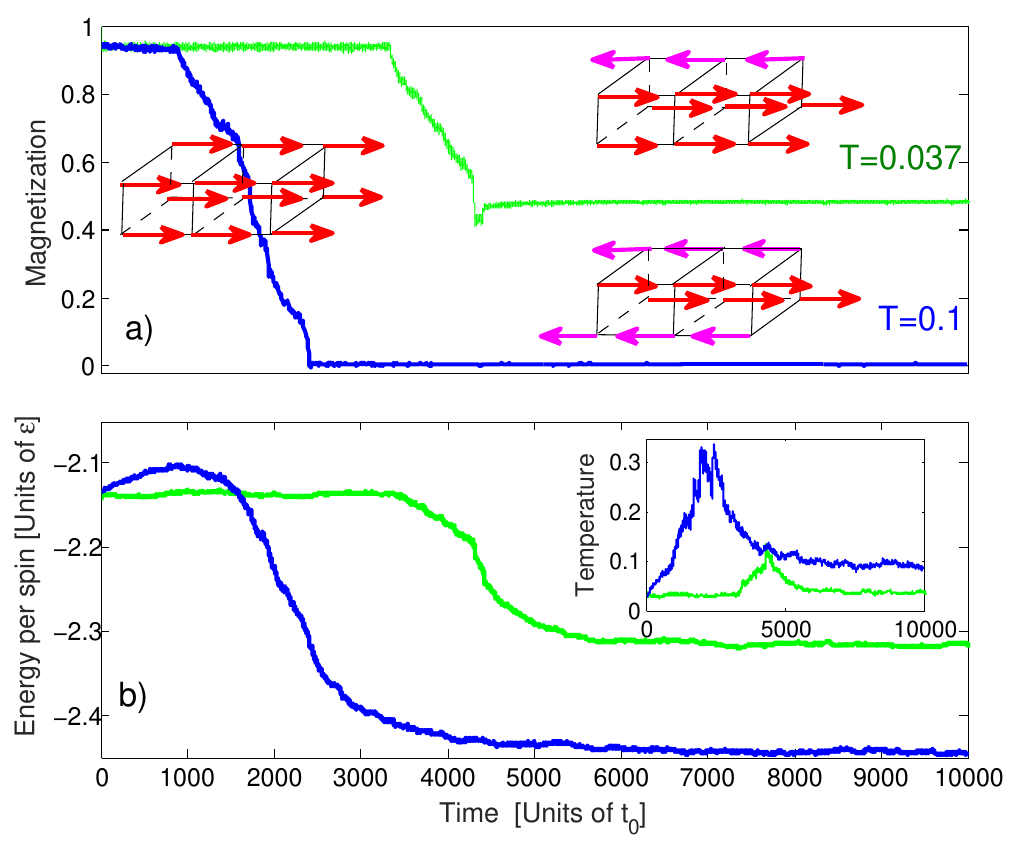}
\caption{Time evolution of magnetization, energy and temperature of initially magnetized dipolar systems in the
canonical ensemble. After the contact with a thermal bath having a temperature $T=0.0375$, corresponding to microcanonical ferromagnetic states, the system transforms into a partially antiferromagnetic state shown as green (light grey) curves. For a larger thermal bath temperature, $T=0.1$, the system goes directly to an antiferromagnetic state displayed as blue (dark grey) curves in the figure.} 
\label{damag}
\end{figure}

In the canonical ensemble, we examine the dipolar spin system by applying a thermal
bath according to Eqs.~\eqref{LLG} and \eqref{bath}. Excluding the temperature range in which
antiferromagnetic and magnetized states coexist, the caloric curves agree with those
of the microcanonical ensemble (see Fig. \ref{ET}), revealing ensemble equivalence.From the same figure one can see that the microcanonical temperature for which magnetized states could be realized in the microcanonical ensemble spans in the range  $0.035<T_\mu<0.04$. 
If one applies the thermal bath with larger temperatures, e.g. $T=0.1$ (see Fig. \ref{damag} blue curves), one observes that magnetized states become quasi-stationary and are destabilized after short time. If instead we attempt to thermalize a magnetized state with the bath temperature in the above range of temperatures for which ferromagnetic states are present in the 
microcanonical ensemble, one observes the emergence of partially antiferromagnetic
states. For example, for the bath temperature $T=0.0375$ the time evolution of the system is shown in Fig.~\ref{damag} with green curves.
\begin{figure}[t]
\includegraphics[scale=.48]{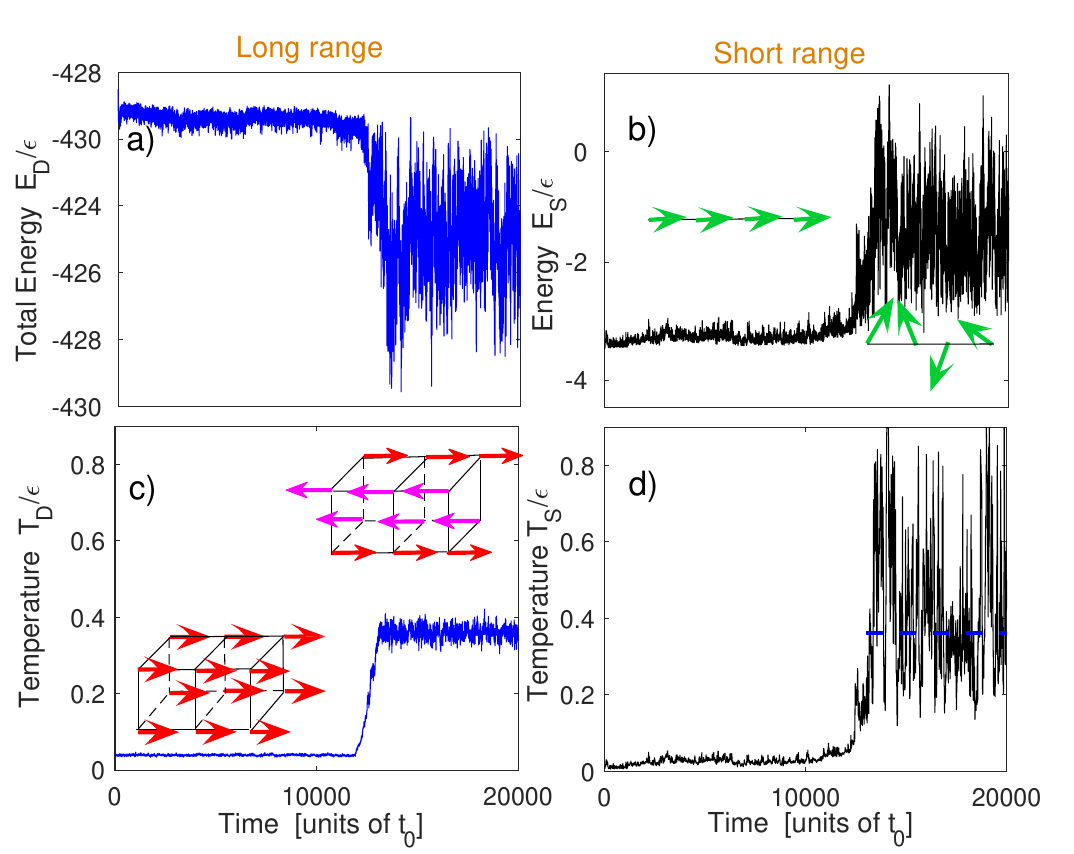}
\caption{Time evolution of temperature and energy for both the dipolar
system and the spin chain during the transition between microcanonical
ferromagnetic state and  antiferromagnetic states.
Both systems are initially prepared in a ferromagnetic state. Panels a) and
b) show the evolution of the energies, while c) and d) display
the evolution of the temperature for the dipolar system and for the
spin chain, respectively. The arrows in b) show schematically the initial
and final distributions of spins in the spin chain, while those in c) represent the transition
from ferromagnetic to antiferromagnetic ordering of the dipolar system. 
The horizontal dashed line in d) shows the average
final temperature of the spin chain, which coincides with the
temperature of the dipolar system in the final state.} 
\label{chainsistem}
\end{figure}

We have also performed numerical experiments in order to analyze
in more detail the transition with negative specific heat in the microcanonical
ensemble. We start the simulation by preparing the dipolar
system with magnetization $\sim 0.94$ and microcanonical temperature $0.035$,
while a spin-chain with 6 spins is prepared in the ground state at zero temperature.
In Fig.~\ref{chainsistem}, we show the time evolution of both the energy and the
temperature for the two systems. For the dipolar system, one clearly
observes a decrease of energy corresponding to an increase of temperature,
while for the spin chain both energy and temperature increase. The
final temperature for the two systems is the same, equal to 0.38 in units of $\epsilon$. 

\begin{figure}[t]
\includegraphics[scale=.45]{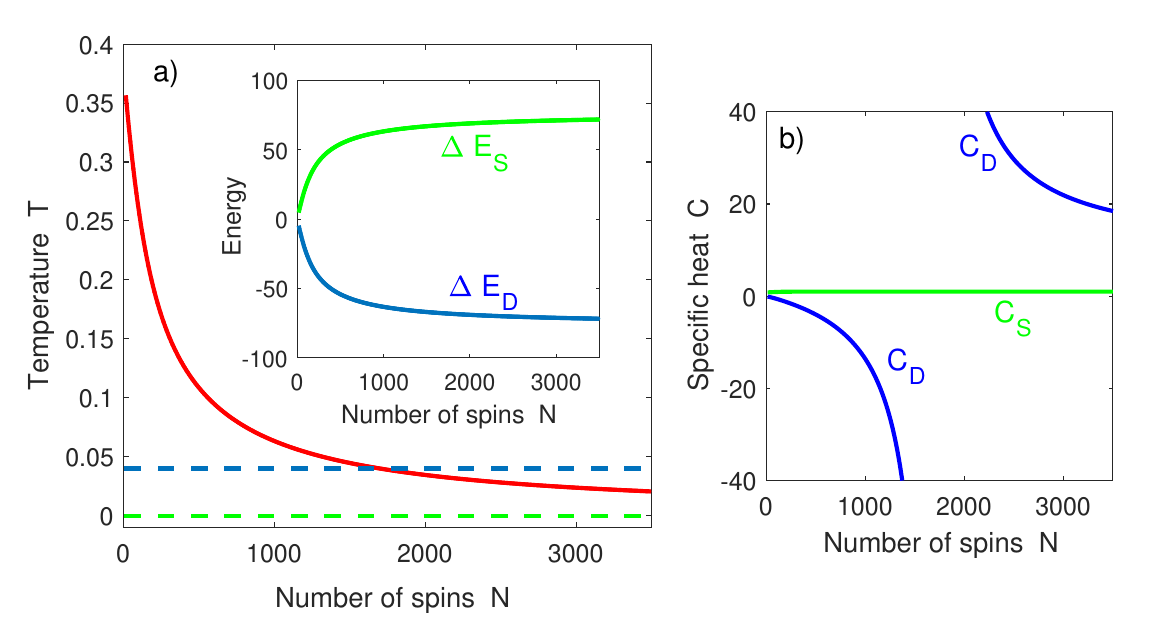}
\caption{Panel a): Final microcanonical temperature (red solid line) reached by the dipolar system after the contact 
with the spin chain as a function of number of spins $N$ in the spin chain. The curve is obtained using energy
conservation, given by relation~\eqref{eq}. The dashed blue (upper) and green (lower) horizontal lines display the initial temperatures of 
the dipolar system and the spin chain, respectively. The inset shows the respective energy gain and loss. 
Panel b): specific heat values $C_D$ and $C_S$ of the dipolar system and spin chain versus the number of spins~$N$ 
in the chain.} 
\label{CVfig}
\end{figure}

The underlying physics of this transition is clearly seen from the
caloric curves presented in Fig.~\ref{ET}. Indeed, the initial  microcanonical
temperature of the dipolar needle is higher than the spin chain
temperature $T_D>T_S$. Therefore, since the energy should flow from
hot to cold, the energy of the dipolar system unavoidably
decreases, which determines the transition to the antiferromagnetic state.
The temperature of the dipolar system increases only if the spin chain is small enough 
because it is not capable of getting a large amount of energy from the dipolar system. 
The final temperature can be heuristically calculated from the total
energy conservation
\begin{equation}
E_D^0+N{\cal E}_S(T_\mu=0)=E_D(T_\mu)+N{\cal E}_S(T_\mu), 
\label{eq}
\end{equation}
where $E_D^0$ is the energy of the magnetized state of the dipolar needle, $N$ is the number of spins in 
the spin chain, ${\cal E}_S(T_\mu)$ is derived by inverting formula~\eqref{lang} and 
$E_D(T_\mu)$ is obtained from the caloric curve for the antiferromagnetic state of the dipolar needle presented in
Fig.~\ref{ET}. When increasing the number of spins $N$ of the spin chain, the temperature 
difference $\Delta T_S$ is even larger providing a larger (proportional to $N$) energy exchange 
between the two systems. As a result, negative specific heat increases in modulus when increasing the number of spins in the spin chain, thus it depends on the size of the spin chain in contact with
the dipolar system. We have analyzed this nontrivial behavior on the basis of relation~\eqref{eq}; see Fig.~\ref{CVfig}. Specific heat values per spin are calculated as $C_D=\Delta E_D/(200\Delta T_D)$ and  $C_S=\Delta E_S/(N\Delta T_S)$.
We then keep the number of spins in the dipolar system unchanged to $200=2\times 2 \times 50$ and vary the number 
$N$ of spins in the chain. Figure~\ref{CVfig} shows that the specific heat of the dipolar system $C_D$ increases 
with $N$ in modulus and stays negative up to values of $N\approx 1500$. For larger values of $N$,  the specific heat
of the dipolar system is positive. As expected, the values of the specific heat of the spin chain, $C_S$, 
stay constant when varying the number of spins in the chain. 

In conclusion, we have shown the presence of negative specific heat in dipolarly coupled three-dimensional spin systems in the microcanonical ensemble. We interpret such a curious thermal behavior as a transition process between different caloric curves induced by the interaction of the bulk of the system with the spin chain. We have realized the canonical ensemble simulations by employing a clever algorithm that uses the notion of Nos\'e-Hoover thermostat~\cite{term}. The dynamical process is totally different in this latter ensemble and finally leads to a partially magnetized state with no negative specific heat. This result confirms the global picture that ensembles behave differently when forces are marginally long-range. In the thermodynamic limit, when both the transversal and the longitudinal size of the dipolar system are increased keeping aspect ratio large,  the fully magnetised state cannot be observed in the microcanonical ensemble as argued in Ref. \cite{miloshevich}. Instead,  a ferrimagnetic state with magnetisation approximately equal to $1/2$ (similar to the one shown in Fig. \ref{damag}) can be realised. Thus, one still has a coexistence of magnetised and antiferromagnetic states in the microcanonical ensemble and the presence of negative specific heat is unavoidable in the transition process. Experiments with magnetic nanoparticles could reveal for the first time the presence of negative specific heat in macroscopic samples.  

We thank William G. Hoover for suggesting Ref.~\cite{term}. R. Kh. and Th. D. acknowledge financial support of Joint grant from Georgian SRNSF and CNRS, France (Grant No. 04/01). This work was supported by Shota Rustaveli
National Science Foundation of Georgia (SRNSFG) [grant number FR-19-4049].


\begin{thebibliography}{abc}
\bibitem{bookLRI}
A. Campa, T. Dauxois, D. Fanelli, S. Ruffo, {\it Physics of Long-Range Interactions}, Oxford University Press (2014).
\bibitem{levin} Y. Levin, R. Pakter, F.B. Rizzato, T.N. Teles, F.P.C. Benetti, Nonequilibrium statistical mechanics of systems with long-range interactions, Phys. Reports, {\bf 535}, 1 (2014).
\bibitem{antonov} V.A. Antonov, Solution of the problem of stability of stellar system Emden’s density law and the spherical distribution of velocities, Vestnik Leningradskogo Universiteta {\bf 7}, 135 (1962).
\bibitem{lynden} D. Lynden-Bell, R. Wood, The gravo-thermal catastrophe in isothermal spheres and the onset of red-giant structure for stellar systems, Monthly Notices of the Royal Astronomical Society {\bf 138},  495 (1968).
\bibitem{thirring} W.Thirring, Systems with Negative Specific Heat, Z. Physik {\bf 235}, 339 (1970).
\bibitem{solid} M. Schmidt, R. Kusche, T. Hippler, J. Donges, W. Kronmüller, B.von Issendorff, H. Haberland, Negative Heat Capacity for a Cluster of 147 Sodium Atoms, Physical Review Letters {\bf 86}, 1191 (2001).
\bibitem{gas} F. Gobet, B. Farizon, M. Farizon, M. J. Gaillard, J. P. Buchet, M. Carré, P. Scheier, and T. D. M\"ark, Direct Experimental Evidence for a Negative Heat Capacity in the Liquid-to-Gas Phase Transition in Hydrogen Cluster Ions: Backbending of the Caloric Curve, Phys. Rev. Lett. {\bf 89}, 183403 (2002).
\bibitem{bouchet} F. Bouchet, A. Venaille, Statistical mechanics of two-dimensional and geophysical flows, Physics Reports  {\bf 515}, 227–295 (2012).
\bibitem{campa1} A. Campa, T. Dauxois and  S. Ruffo, Statistical mechanics and dynamics of solvable models with long-range interactions, Physics Reports  {\bf 480}, 57-159 (2009).
\bibitem{PRL1} J. Barr{\'e}, D. Mukamel, S. Ruffo, Inequivalence of Ensembles in a System with Long-Range Interactions, Physical Review Letters  {\bf 87}, 030601 (2001).
\bibitem{mukamel} A. Campa, R. Khomeriki, D. Mukamel, S. Ruffo, Long-range effects in layered spin structures, Phys. Rev. B, {\bf 76}, 064415 (2007).
\bibitem{miloshevich} G. Miloshevich, T. Dauxois, R. Khomeriki, S. Ruffo, Dipolar needles in the microcanonical ensemble: evidence of spontaneous magnetization and ergodicity breaking,  Europhysics Letters {\bf 104}, 17011 (2013).
\bibitem{griffiths} S. Banerjee, R. B. Griffiths, M. Widom, Thermodynamic Limit for Dipolar Media, J. Stat. Phys., {\bf 93}, 109 (1998).
\bibitem{term}D. Kusnezov and A. Bulgac, Canonical Ensembles from Chaos, Annals of Physics {\bf 214}, 180-218 (1992)
\bibitem{cobalt} M. Varon, M. Beleggia, T. Kasama, R. J. Harrison, R. E. Dunin-Borkowski, V. F. Puntes, C. Frandsen, Dipolar Magnetism in Ordered and Disordered Low-Dimensional Nanoparticle Assemblies, Scientific Reports {\bf 3},  1234 (2013).
\bibitem{mukruf} D. Mukamel, S. Ruffo, N. Schreiber, Breaking of Ergodicity and Long Relaxation Times in Systems with Long-Range Interactions, Physical Review Letters {\bf 95}, 240604 (2005).
\bibitem{rugh} H. H. Rugh, Dynamical Approach to Temperature, Physical Review Letters {\bf 78}, 772 (1997).
\bibitem{gut} C. Jara, F. Gonz\'alez-Cataldo, S. Davis and G.
Guti\'errez, Temperature estimators in computer simulation, Journal of Physics: Conference Series {\bf 720}, 012003 (2016).
\bibitem{wira} W.B. Nurdin, K.-D. Schotte, Dynamical temperature for spin systems, Phys. Rev. E. {\bf 61}, 3579 (2000).
\bibitem{Nose} S. Nos\'e, A unified formulation of the constant temperature molecular-dynamics methods, Journal of Chemical Physics {\bf 81} 511 (1984); W. G. Hoover, Canonical dynamics: equilibrium phase-space distributions, Phys. Rev. A. {\bf 31}, 1695 (1985).
\bibitem{symplectic} As it has been proposed in~\cite{yoshida}, we split our Hamiltonian in a sum of several parts, e.g. ${\cal H}={\cal A}+{\cal B}$, where each of them (i.e. ${\cal A}$ and ${\cal B}$) alone is exactly integrable. Then, the evolution operator can be approximated at second order in the time step $\tau$ as $\exp(\tau{\cal H})=\exp(\tau{\cal A}/2)\exp(\tau{\cal B})\exp(\tau{\cal A}/2)+{\cal O}(\tau^3)$. Since the evolution of any arbitrary initial condition under the operators 
$\exp(\tau{\cal A}/2)$ and $\exp(\tau{\cal B})$ is exactly known, we can determine the total evolution process at second
order in $\tau$.
\bibitem{yoshida} H. Yoshida, Construction of higher order symplectic integrators, Phys. Lett. A. {\bf 150}, 262 (1990).

 

\end{thebibliography}
\end{document}